\documentclass[twocolumn,aps,prl,groupedaddress]{revtex4}
\usepackage{amssymb}

\usepackage{graphicx}
\usepackage{dcolumn}
\usepackage{bm}
\usepackage{pifont}

\begin{document}

\title{Critical behavior of a strongly-interacting 2D electron system}
\author{A. Mokashi$^{(a)}$, S. Li$^{(b)}$, B. Wen$^{(b)}$, S.~V. Kravchenko$^{(a)}$, , A.~A. Shashkin$^{(c)}$, V.~T. Dolgopolov$^{(c)}$, and M.~P. Sarachik$^{(b)}$}
\affiliation{$^{(a)}$Physics Department, Northeastern University, Boston, Massachusetts 02115, USA}
\affiliation{$^{(b)}$Physics Department, City College of the City University of New York, New York, New York 10031, USA}
\affiliation{$^{(c)}$Institute of Solid State Physics, Chernogolovka, Moscow District 142432, Russia}
\begin{abstract}
With decreasing density $n_s$ the thermopower $S$ of a low-disorder 2D electron system in silicon is found to exhibit a sharp increase by more than an order of magnitude, tending to a divergence at a finite, disorder-independent density $n_t$ consistent with the critical form $(-T/S) \propto
(n_s-n_t)^x$ with $x=1.0\pm 0.1$ ($T$ is the temperature). Our
results provide clear evidence for an interaction-induced transition to a new phase at low density in a strongly-interacting 2D electron system.
\end{abstract}
\maketitle

The behavior of strongly-interacting electrons in two dimensions
(2D) is a forefront area of condensed matter physics in which
theoretical methods are still poorly developed and new
experimental results are of great interest. Consistent with Fermi liquid theory at high electron densities \cite{landau57}, these 2D systems are expected to undergo one or more transitions to spatially and/or spin-ordered phases as the density is decreased, ultimately forming a Wigner crystal in the dilute, strongly-interacting limit \cite{wigner34,stoner47,chaplik72,tanatar89,attaccalite02}. The
interaction strength is characterized by the ratio of the Coulomb energy to the Fermi energy, determined by the dimensionless parameter, $r_s= 1/(\pi n_sa_B^2)^{1/2}$ (here $n_s$ is the areal density of electrons, $a_B =
\varepsilon\hbar^2/m_be^2$, and $\varepsilon$, $e$, and $m_b$ are
the dielectric constant, the absolute value of electron charge,
and the band mass, respectively); the parameter $r_s$ is
proportional to $n_s^{-1/2}$ and increases with decreasing
electron density, reaching values in excess of $r_s\gtrsim10$ in
systems investigated experimentally to date. Particularly strong
many-body effects have been observed in silicon
metal-oxide-semiconductor field-effect transistors.

In this Letter we report that the thermopower of a low-disorder 2D electron system in silicon exhibits critical behavior with decreasing electron density, tending toward a divergence at a well-defined disorder-independent density $n_t$.  Our results provide clear evidence for an interaction-induced transition to a new phase at low density which may be a precursor phase, or a direct transition to the long sought-after Wigner solid.

\begin{figure}
\scalebox{0.47}{\includegraphics{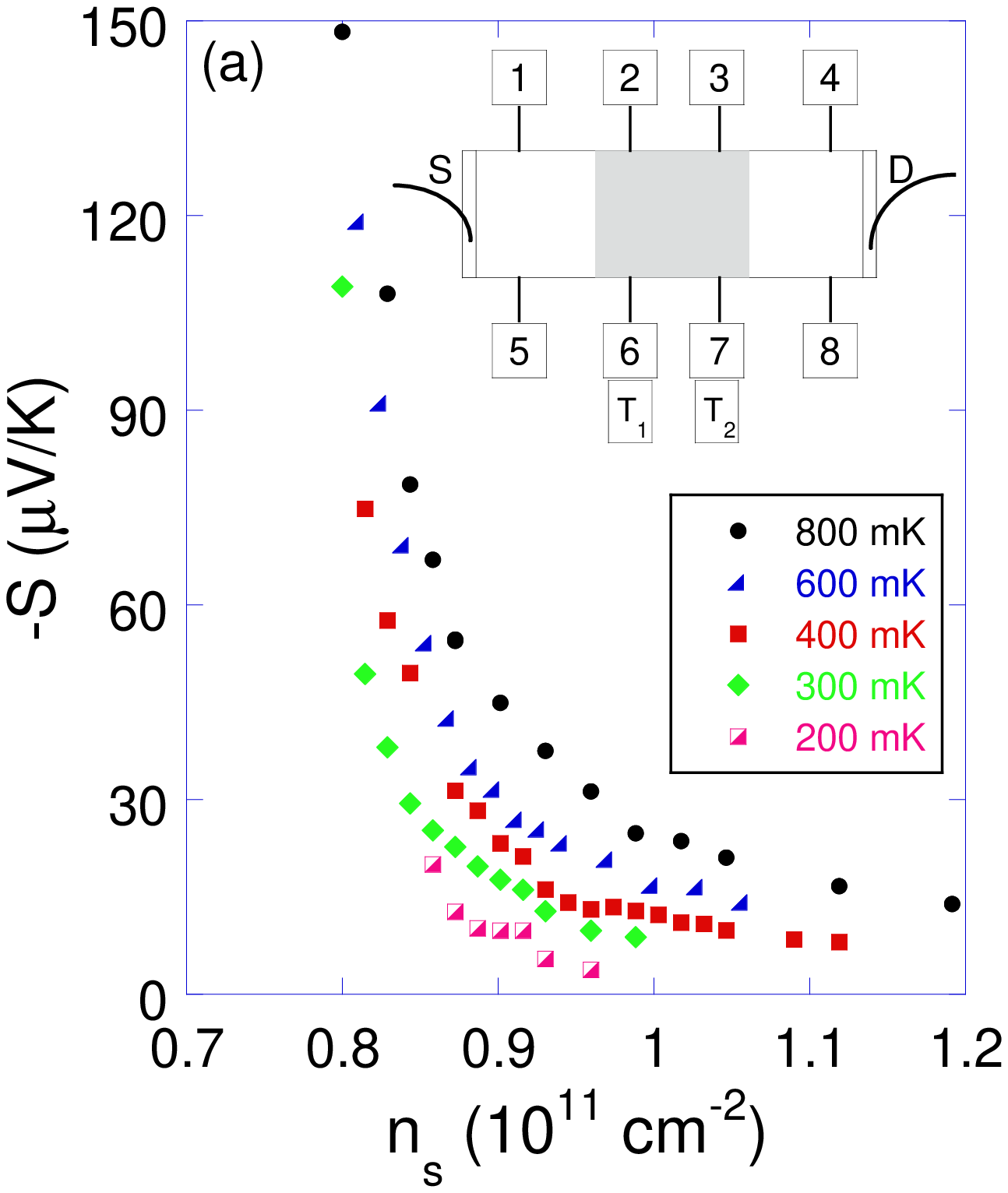}}
\scalebox{0.47}{\includegraphics{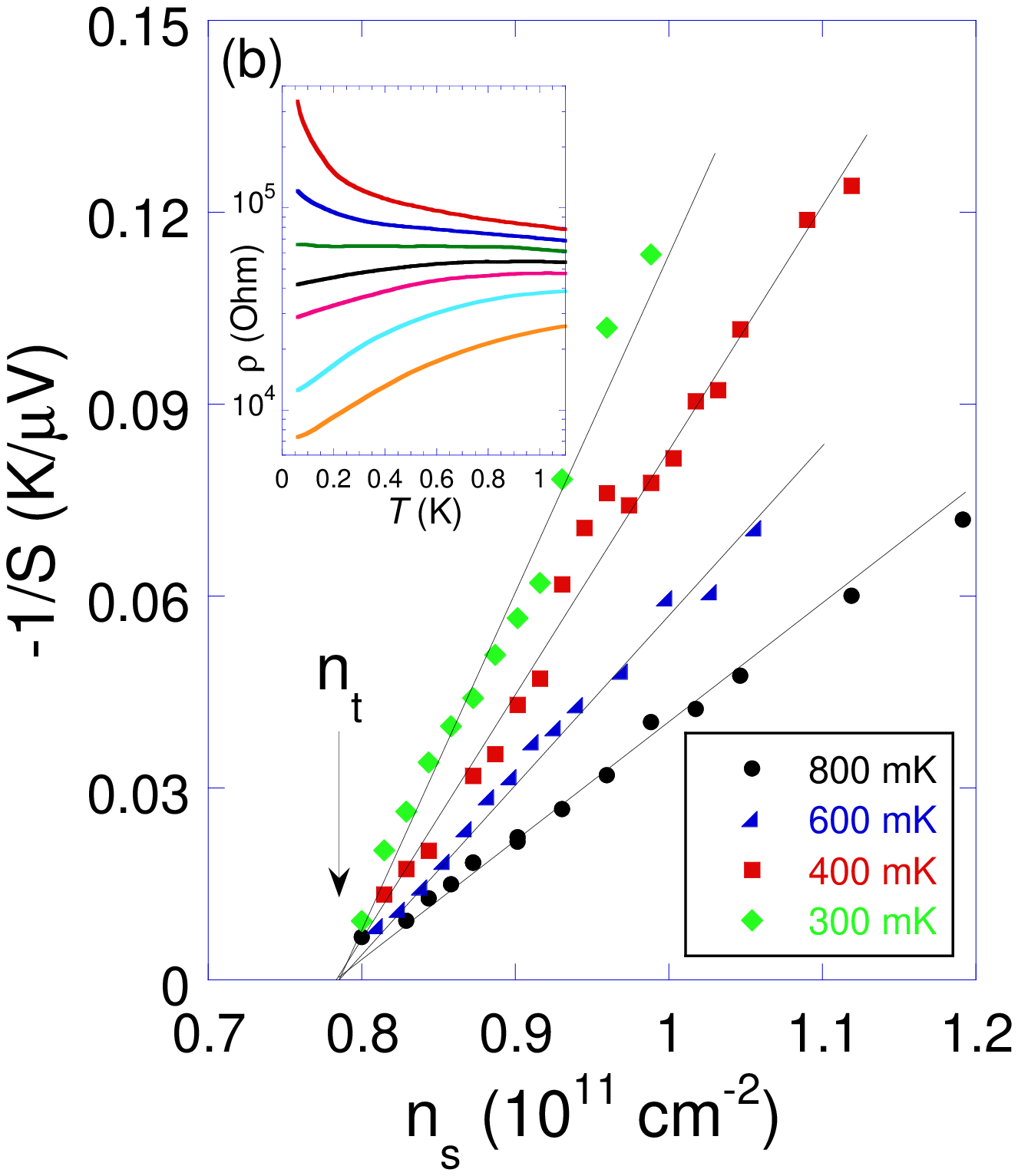}}
\caption{\label{fig1} (Color online) (a) Thermoelectric power, $S$, as a function
of electron density $n_s$ at different temperatures. Many data
points are omitted for clarity. The inset is a schematic view of
the sample. The contacts include four pairs of potential probes,
source, and drain; the main part of the sample is shaded. The
thermometers $T_1$ and $T_2$ measure the temperature of the contacts. (b) The inverse thermopower as a function of electron density at different
temperatures. The solid lines denote linear fits to the data and
extrapolate to zero at a density $n_t$.  The inset shows the resistivity as a function of temperature for electron densities (top to bottom): $0.768, 0.783, 0.798, 0.813, 0.828, 0.870$, and $0.914 \times 10^{11}$ cm$^{-2}$.}
\end{figure}

The thermopower is defined as the ratio of the thermoelectric
voltage to the temperature difference, $S=-\Delta V/\Delta T$.
Measurements were made in a sample-in-vacuum Oxford dilution
refrigerator with a base temperature of $\approx 30$~mK on
(100)-silicon metal-oxide-semiconductor field-effect transistors similar to those previously used in
Ref.~\cite{heemskerk98}. The advantage of these samples is a very
low contact resistance (in ``conventional'' silicon samples, high
contact resistance becomes the main experimental obstacle in the
low-density low-temperature limit). To minimize contact
resistance, thin gaps in the gate metallization have been
introduced, which allows for maintaining high electron density
near the contacts regardless of its value in the main part of the
sample. The electron density was controlled by applying a positive dc voltage to the gate relative to the contacts; the oxide thickness was $150$ nm.  Samples were used with a Hall bar geometry of width
50~$\mu$m and distance 120~$\mu$m between the central potential
probes, and measurements of the thermoelectric voltage were
obtained in the main part of the sample (shaded in the inset to
Fig.~\ref{fig1}(a)). A Hall contact pair, either 1-5 or 4-8, was
employed as a heater: the 2D electrons were locally heated by
passing an ac current at a low frequency $f$ through either pair.
Both the source and drain contacts were thermally anchored. In such an
arrangement it was possible to reverse the direction of the
temperature gradient induced in the central region of the sample.
The temperatures of the central probes were determined using two
thermometers glued to the metallic pads on the sample holder connected by metallic wires to the contacts on the sample; temperature gradients between contacts reached 1-5 mK over the distance. The measured temperatures were independent of the electron density in the central region, indicating that the heat flowed from the heater to the anchor through the lattice, so that our experiment is similar to a standard set-up for thermopower measurements.  The average temperature determined by the thermometers was checked to correspond to the average electron temperature in the central region measured using the calibrated sample resistivity.  The temperature difference between the pairs of contacts 6, 7, and the source or drain along the thermal path from the heater to the anchor was monitored and found to be proportional to the distance between the contacts, as expected.  Constantan or superconducting wiring was employed to minimize heat leaks from the sample. Possible rf pick-up was carefully suppressed, and the thermoelectric voltage was measured using a low-noise low-offset
LI-75A preamplifier and a lock-in amplifier in the $2f$ mode in
the frequency range 0.01--0.1~Hz. The sample resistance was
measured by a standard 4-terminal technique at a frequency 0.4~Hz.
Excitation currents were kept sufficiently small (0.1--1~nA) to
ensure that measurements were taken in the linear regime. The
results shown in this Letter were obtained on a sample with a peak
electron mobility close to 3~m$^2$/Vs at $T=0.1$~K.

\begin{figure}
\scalebox{0.41}{\includegraphics{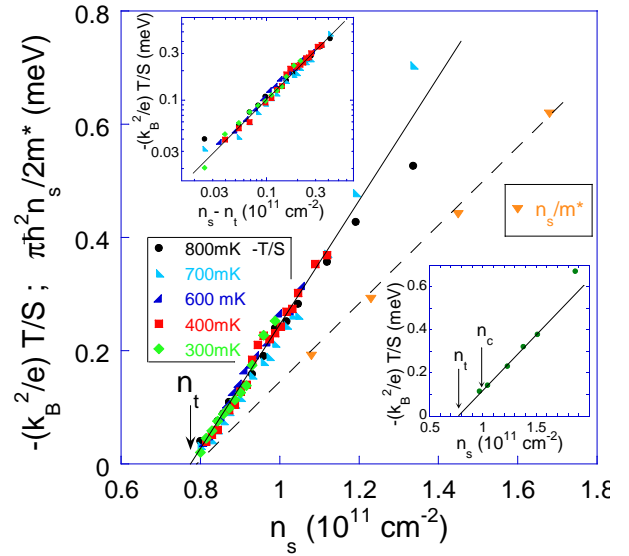}}
\caption{\label{fig2} (Color online) $(-T/S)$ versus electron density $n_s$ for
different temperatures. The solid line is a linear fit which
extrapolates to zero at $n_t$. Also shown is the effective mass
$m^*$ obtained for the same samples by different measurements
\cite{shashkin02}. The dashed line is a linear fit. Inset, upper
left-hand corner: log-log plot of $(-T/S)$ versus ($n_s-n_t)$,
demonstrating power law approach to the critical density $n_t$;
Inset, lower right-hand corner: $(-T/S)$ versus density at
$T=0.3$~K for a highly-disordered 2D electron system in silicon
\cite{fletcher01}. The linear fit (solid line) extrapolates to
zero at the same density $n_t$. The position of the density $n_c$
for the metal-insulator transition was estimated to be $0.99\pm
0.02\times 10^{11}$~cm$^{-2}$.}
\end{figure}

\begin{figure}
\scalebox{0.47}{\includegraphics{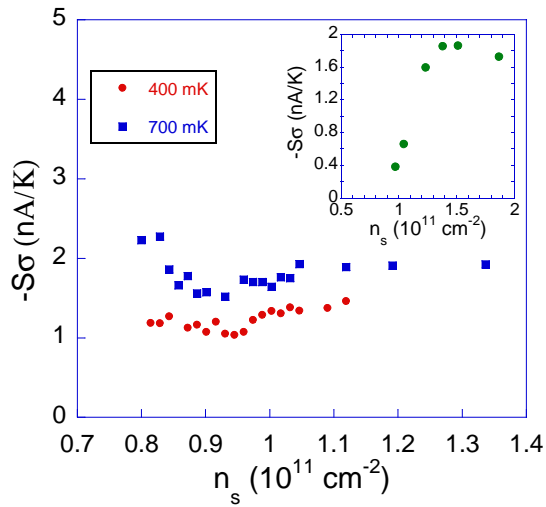}}
\caption{\label{fig3} (Color online) The product $(-S\sigma)$ that determines the
thermoelectric current plotted as a function of electron density
$n_s$ at different temperatures. Inset: $(-S\sigma)$ versus
electron density at $T=0.3$~K for a highly-disordered 2D electron
system in silicon \cite{fletcher01}. The density $n_c$ for the
metal-insulator transition in this high-disorder sample is shown
in the lower right-hand inset of Fig.~\ref{fig2}.}
\end{figure}

Our experimental results are shown in Figs.~\ref{fig1} through
\ref{fig3}. Figure~\ref{fig1}(a) shows data for the thermopower as
a function of $n_s$ at different temperatures.
$(-S)$ increases strongly with decreasing electron density and
becomes larger as the temperature is increased. The divergent
behavior of the thermopower is evident when plotted as the inverse
quantity $(-1/S)$ versus electron density in Fig.~\ref{fig1}(b).

Figure~\ref{fig2} shows $(-T/S)$ plotted as a function of $n_s$.
The data collapse onto a single curve demonstrating that the
thermopower $S$ is a linear function of temperature. In turn, the
ratio $(-T/S)$ is a function of electron density $n_s$ of the
form:
\begin{equation}
(-T/S) \propto (n_s-n_t)^x.\label{eq1}
\end{equation}
Fits to this expression indicate that the thermopower diverges
with decreasing electron density with a critical exponent $x=1.0
\pm 0.1$ at a density $n_t = 7.8 \pm 0.1 \times
10^{10}$~cm$^{-2}$ that is close to (or the same as) the density
for the metal-insultator transition $n_c \approx 8 \times
10^{10}$~cm$^{-2}$, obtained from resistivity measurements in this
low-disorder electron system (see the inset to Fig.~\ref{fig1}(b)). The log-log plot shown in the inset
(upper left-hand corner) of Fig. \ref{fig2} demonstrates the
critical, power law, behavior of the thermopower.

In Fig.~\ref{fig3} we show the product $(-S\sigma)$ that
determines the thermoelectric current $j=-S\sigma\nabla T$ as a
function of electron density at two different temperatures (here
$\sigma$ is the conductivity). $(-S\sigma)$ is approximately
constant in the critical region, i.e., $(1/S)$ is proportional to
$\sigma$ in the low-disorder 2D electron system. Within the
relaxation time approximation, one expects the thermopower $S$
to depend only weakly on scattering, while the scattering should
play a major role in determining the conductivity. That
($S\sigma$) is constant signals that disorder is not the origin of
the critical behavior in our samples, which derives instead from
strong electron-electron interactions. The fact that the behavior
shown in Fig.~\ref{fig2} continues smoothly down to the lowest
electron densities achieved confirms that the disorder effects
that might cause deviations are minor.

Confirmation is provided by comparison with earlier data obtained
by Fletcher {\it et al.} \cite{fletcher01} in a silicon sample
with a high level of disorder, as indicated by the appreciably
higher density $n_c$ for the resistively determined
metal-insulator transition. A replot of the thermopower taken from
Ref.~\cite{fletcher01}, shown in the lower right-hand inset of
Fig.~\ref{fig2}, demonstrates that $(- T/S)$ measured well above
the critical point extrapolates to the same density $n_t$.
However, in contrast with our data, $(-S\sigma$) for the
higher-disorder silicon samples tends to zero at the
higher-density transition point $n_c$ (see inset to
Fig.~\ref{fig3}) due to a rapidly decreasing conductivity $\sigma$
for $n_s<n_c$. Thus, while the resistive transition $n_c$ varies
with disorder, the divergence of the thermopower occurs at a
density $n_t$ that is independent of disorder \cite{remark}. This
indicates clearly that the transitions in low- and high-disorder
silicon derive from different sources: whereas in
highly-disordered 2D electron systems the conductivity tends to
zero due to disorder, in the clean 2D electron system the drop of
the conductivity occurs at the transition driven by
electron-electron interactions \cite{shashkin02}.

Based on Fermi liquid theory, Dolgopolov and Gold
\cite{dolgopolov11,gold11} recently obtained the following
expression for the diffusion thermopower of strongly interacting
2D electrons in the low-temperature regime:
\begin{equation}
S=-\alpha\frac{2\pi k_B^2mT}{3e\hbar^2n_s},\label{eq2}
\end{equation}
where $k_B$ is Boltzmann's constant and $m$ is the effective mass.
This expression, which resembles the well-known Mott relation for
non-interacting electrons, was shown to hold for the
strongly-interacting case provided one includes the parameter
$\alpha$ that depends on both the disorder
\cite{fletcher97,faniel07,goswami09} and interaction strength
\cite{dolgopolov11,gold11}. The dependence of $\alpha$ on electron
density is rather weak, and the main effect of electron-electron
interactions is to suppress the thermopower $S$.

Note that we have found $S \propto T$, as expected for the
diffusion thermopower. This indicates that the phonon drag
contribution is small in the temperature range of our experiments,
and our measurements yield the contribution of interest, namely,
the diffusion thermopower.

The measured $(-T/S)$, shown in Fig.~\ref{fig2}, decreases
linearly with decreasing electron density, extrapolating to zero
at $n_t$. According to Eq.~(\ref{eq2}), $(-T/S)$ is proportional
to $(n_s/m)$, indicating a strong increase of the mass by more
than an order of magnitude.  Our results thus imply a divergence
of the electron mass at the density $n_t$: $m\propto
n_s/(n_s-n_t)$ --- behavior that is typical in the vicinity of an
interaction-induced phase transition.

It is interesting to compare these results with the effective mass
$m^*$ obtained earlier for the same samples, where $m^*$ and the
$g$-factor were determined by combining measurements of the slope
of the conductivity versus temperature with measurements of the
parallel magnetic field $B^*$ for full spin polarization
\cite{shashkin02}. As seen in Fig.~\ref{fig2}, the two data sets
display similar behavior. However, the thermopower data do not
yield the absolute value of $m$ because of uncertainty in the
coefficient $\alpha$ in Eq.~(\ref{eq2}). The value of $m$ can be
extracted from the thermopower data by requiring that the two data
sets in Fig.~\ref{fig2} correspond to the same value of mass in
the range of electron densities where they overlap. Determined
from the ratio of the slopes, this yields a coefficient
$\alpha\approx 0.18$. The corresponding mass enhancement in the
critical region reaches $m/m_b\approx 25$ at $n_s\approx 8.2\times
10^{10}$~cm$^{-2}$, where the band mass $m_b=0.19m_e$ and $m_e$ is
the free electron mass. The mass $m\approx 5m_e$ exceeds by far
the values of the effective mass obtained from previous
experiments on the 2D electron system in silicon as well as other
2D electron systems.

It is important to note that the current experiment includes data
for electron densities that are much closer to the critical point
than the earlier measurements, and reports much larger enhancement
of the effective mass for reasons explained below.

The Zeeman field $B^*$ required to fully polarize the spins and
the thermopower measurements both imply a large enhancement of the
effective mass \cite{remark1}. However, the two experiments
measure different effective masses: the thermopower gives a
measure of the mass at the Fermi level, while $B^*$ measures the
mass related to the bandwidth, which is the Fermi energy counted
from the band bottom. In other words, while the thermopower, as
well as the conductivity, are sensitive to the low energy
excitations within an energy range $\sim k_BT$ near the Fermi
energy, the Zeeman field $B^*$ for full spin polarization is a
measure of the bandwidth and is sensitive to the behavior of all
states including those relatively far from the Fermi energy.

For $n_s\geq 10^{11}$~cm$^{-2}$, the mass was found to be
essentially the same \cite{kravchenko04,shashkin05}, thereby
justifying our determination of $\alpha$. On the other hand, the
behavior is different at the densities reached in our experiment
in the very close vicinity of the critical point $n_t$
($n_s<10^{11}$~cm$^{-2}$), where the bandwidth-related mass was
found to increase by only a factor $\approx 4$. Indeed, we argue
that the bandwidth-related mass does not increase strongly near
$n_t$. If so, the ratio of the spin and cyclotron splittings in
perpendicular magnetic fields would increase considerably with
decreasing electron density so that the spin-up and spin-down
levels should cross whenever this ratio is an integer. One should
then observe a Shubnikov-de~Haas oscillation beating pattern with
decreasing electron density, including several switches between
the oscillation numbers in weak magnetic fields. Instead, the
Shubnikov-de~Haas oscillations in the dilute 2D electron system in
silicon reveal one switch from cyclotron to spin minima (the ratio
of the spin and cyclotron splittings reaches $\approx 1$) as the
electron density is decreased \cite{kravchenko00}, the spin minima
surviving down to $n_s\approx n_c$ and even below \cite{iorio90}.

In effect, while the bandwidth does not decrease appreciably in
the close vicinity of the critical point $n_t$ and the effective
mass obtained from such measurements does not exhibit a true
divergence, the thermopower measurements yield the effective mass
at the Fermi energy, which does indeed diverge.

A divergence of the effective mass has been predicted by a number
of theories: by using Gutzwiller's theory \cite{dolgopolov02}, by using
an analogy with He$^3$ near the onset of Wigner crystallization
\cite{spivak03,spivak04}, by extending the Fermi liquid concept to
the strongly-interacting limit \cite{khodel08}, by solving an
extended Hubbard model using dynamical mean-field theory
\cite{pankov08}, by using a renormalization group analysis for
multi-valley 2D systems \cite{punnoose05}, and by using Monte-Carlo
simulations \cite{marchi09,fleury10}. Some theories predict that
the disorder is important for the mass enhancement
\cite{punnoose05,marchi09,fleury10}. In contrast with most
theories that assume a parabolic spectrum, the authors of
Ref.~\cite{khodel08} stress that there is a clear distinction
between the mass at the Fermi level and the bandwidth-related
mass. In this respect, our conclusions are consistent with the
model of Ref.~\cite{khodel08} in which a flattening at the Fermi
energy in the spectrum leads to a diverging effective mass. This
Fermi liquid-based model implies the existence of an intermediate
phase that precedes Wigner crystallization.

There has been a great deal of debate concerning the
origin of the interesting, enigmatic behavior in these strongly
interacting 2D electron systems. In particular, many have
questioned whether the change of the resistivity from metallic to
insulating temperature dependence signals a phase transition, or
whether it is a crossover. We close by noting that unlike the
resistivity, which displays complex behavior that may not
distinguish between these two scenarios, we have shown that the
thermopower diverges at a well-defined density, providing clear
evidence that this is a transition to a new phase at low
densities. The next challenge is to determine the nature of this
phase.

We gratefully acknowledge discussions with B.~L. Gallagher, S.~A.
Kivelson, and C.~J. Mellor, and technical help from Y. Zhao. This
work was supported by DOE Grant DE-FG02-84ER45153, BSF grant \#
2006375, RFBR, RAS, and the Russian Ministry of Sciences.


\begin{thebibliography}{99}
\bibitem{landau57} L.~D. Landau, Sov.\ Phys.\ JETP\ {\bf 3}, 920 (1957).
\bibitem{wigner34} E. Wigner, Phys.\ Rev.\ {\bf 46}, 1002 (1934).
\bibitem{stoner47} E.~C. Stoner, Rep.\ Prog.\ Phys.\ {\bf 11}, 43 (1947).
\bibitem{chaplik72} A.~V. Chaplik, Sov.\ Phys.\ JETP\ {\bf 35}, 395 (1972).
\bibitem{tanatar89} B. Tanatar and D.~M. Ceperley, Phys.\ Rev.\ B\ {\bf 39}, 5005 (1989).
\bibitem{attaccalite02} C. Attaccalite, S. Moroni, P. Gori-Giorgi, and G.~B. Bachelet, Phys.\ Rev.\ Lett.\ {\bf 88}, 256601 (2002).
\bibitem{heemskerk98} R. Heemskerk and T.~M. Klapwijk, Phys.\ Rev.\ B\ {\bf 58}, R1754 (1998).
\bibitem{fletcher01} R. Fletcher, V.~M. Pudalov, A.~D.~B. Radcliffe, and C. Possanzini, Semicond.\ Sci.\ Technol.\ {\bf 16}, 386 (2001).
\bibitem{remark} Note that a theory for the thermopower near the Anderson transition has not been developed. Our experimental results nevertheless establish the important fact that the thermopower tends toward a divergence at an electron density $n_t$ that is distinctly different from the density for the Anderson transition.
\bibitem{shashkin02} A.~A. Shashkin, S.~V. Kravchenko, V.~T. Dolgopolov, and T.~M. Klapwijk, Phys.\ Rev.\ B\ {\bf 66}, 073303 (2002).
\bibitem{dolgopolov11} V.~T. Dolgopolov, and A. Gold, JETP\ Lett.\ {\bf 94}, 481 (2011).
\bibitem{gold11} A. Gold and V.~T. Dolgopolov, Europhys.\ Lett.\ {\bf 95}, 27007 (2011).
\bibitem{fletcher97} R. Fletcher, V.~M. Pudalov, Y. Feng, M. Tsaousidou, and P.~N. Butcher, Phys.\ Rev.\ B\ {\bf 56}, 12422 (1997).
\bibitem{faniel07} S. Faniel, L. Moldovan, A. Vlad, E. Tutuc, N. Bishop, S. Melinte, M. Shayegan, and V. Bayot, Phys.\ Rev.\ B\ {\bf 76}, 161307(R) (2007).
\bibitem{goswami09} S. Goswami, C. Siegert, M. Baenninger, M. Pepper, I. Farrer, D.~A. Ritchie, and A. Ghosh, Phys.\ Rev.\ Lett.\ {\bf 103}, 026602 (2009).
\bibitem{remark1} Measurements in the parallel-field configuration yield the product of the effective mass and $g$ factor, where the increase was shown to derive from a strong increase of the electron mass while the $g$ factor enhancement is weak \cite{kravchenko04,shashkin05}.
\bibitem{kravchenko04} S.~V. Kravchenko and M.~P. Sarachik, Rep.\ Prog.\ Phys.\ {\bf 67}, 1 (2004).
\bibitem{shashkin05} A.~A. Shashkin, Phys.\ Usp.\ {\bf 48}, 129 (2005).
\bibitem{kravchenko00} S.~V. Kravchenko, A.~A. Shashkin, D.~A. Bloore, and T.~M. Klapwijk, Solid\ State\ Commun.\ {\bf 116}, 495 (2000).
\bibitem{iorio90} M. D'Iorio, V.~M. Pudalov, and S.~G. Semenchinsky, Phys.\ Lett.\ A\ {\bf 150}, 422 (1990).
\bibitem{dolgopolov02} V.~T. Dolgopolov, JETP\ Lett.\ {\bf 76}, 377 (2002).
\bibitem{spivak03} B. Spivak, Phys.\ Rev.\ B\ {\bf 67}, 125205 (2003).
\bibitem{spivak04} B. Spivak and S.~A. Kivelson, Phys.\ Rev.\ B\ {\bf 70}, 155114 (2004).
\bibitem{khodel08} V.~A. Khodel, J.~W. Clark, and M.~V. Zverev, Phys.\ Rev.\ B\ {\bf 78}, 075120 (2008).
\bibitem{pankov08} S. Pankov and V. Dobrosavljevi\'c, Phys.\ Rev.\ B\ {\bf 77}, 085104 (2008).
\bibitem{punnoose05} A. Punnoose and A.~M. Finkelstein, Science\ {\bf 310}, 289 (2005).
\bibitem{marchi09} M. Marchi, S. De Palo, S. Moroni, and G. Senatore, Phys.\ Rev.\ B\ {\bf 80}, 035103 (2009).
\bibitem{fleury10} G. Fleury and X. Waintal, Phys.\ Rev.\ B\ {\bf 81}, 165117 (2010).
\end{thebibliography}
\end{document}